# Formalized Identification Of Key Factors In Safety-Relevant Failure Scenarios

Tim Maurice Julitz[a], Nadine Schlüter[a], Manuel Löwer[a]

[a]*Departement of Product Safety and Quality Engineering, University of Wuppertal, Germany*

**Abstract**

This research article presents a methodical data-based approach to systematically identify key factors in safety-related failure scenarios, with a focus on complex product-environmental systems in the era of Industry 4.0. The study addresses the uncertainty arising from the growing complexity of modern products. The method uses scenario analysis and focuses on failure analysis within technical product development. The approach involves a derivation of influencing factors based on information from failure databases. The failures described here are documented individually in failure sequence diagrams and then related to each other in a relationship matrix. This creates a network of possible failure scenarios from individual failure cases that can be used in product development. To illustrate the application of the methodology, a case study of 41 Rapex safety alerts for a hair dryer is presented. The failure sequence diagrams and influencing factor relationship matrices show 46 influencing factors that lead to safety-related failures. The predominant harm is burns and electric shocks, which are highlighted by the active and passive sum diagrams. The research demonstrates a robust method for identifying key factors in safety-related failure scenarios using information from failure databases. The methodology provides valuable insights into product development and emphasizes the frequency of influencing factors and their interconnectedness.

## 1. Introduction

### 1.1. Challenges in developing complex products

Modern product development is confronted with novel challenges in light of emerging trends like digitalization, automation, and connectivity, which collectively can be summed up as Industry 4.0. Phenomena such as the Internet of Things or smart factories are increasingly blurring the boundaries of systems and making it more difficult to distinguish them from the environment. This is characterized by a growing variety of products and increased system dynamics, which can be described as complexity, leading to an uncertainty: Products are becoming so complex that the extent of their impact on the usage phase cannot be fully predicted at the time of development.

The observation that the number of transistors on microchips and their performance doubles every 2 years led to the formulation of Moore's Law (Moore, 1998). Today, Intel confirms this trend and assumes that 1 trillion transistors will be installed in a semiconductor package by 2030 (Intel, 2023). Complex systems are influenced by uncertainty which arises through product variety, market fluctuations and their effects. Increased variety leads to the creation of additional information and opens up possibilities for unforeseen or unusual impacts of a product, process or system. This increases the amount of data and expertise required to manage the resulting consequences (ElMaraghy et al., 2012). The automotive sector provides an illustrative example of this. A single modern vehicle can integrate up to 150 electronic control units (ECUs) (Martínez-Cruz et al., 2021). The complexity of these ECUs is further increased by the rising demand for functions and data transfer, e.g. through automated driving (Wartzek et al., 2020). In the context of a system of systems, this results in a large number of possible interactions and interdependencies that cannot be fully predicted and controlled in the development phase of the product. Conventional methods of static failure analysis reach their limits here. Engineers need new methods to deal with this uncertainty in order to react dynamically to unforeseen problems. In this context, the continuous improvement of the product throughout its lifecycle is becoming more important. This also results in a release problem. When has the necessary maturity for the start of series production been reached if the holistic extent of its impact has not yet been determined?

**1.2. Managing Complexity with Scenario Analysis**

In order to manage complexity in product development, the potential interactions between systems, subsystems and the environment must be systematically analyzed. Scenario analysis offers a method for this, since uncertainty and complexity of product systems can be considered by scenario-based approaches (Amer et al., 2013). Scenarios are hypothetical sequences of events which allow the evaluation of a possible chain of causal events (Kahn and Wiener, 1967). An event is an occurrence at a point in time (ISO, 2022). The sequence of events begins with causes and ends with a result. The scenario analysis process is well-known in the context of strategic business or product planning, see Figure 1.

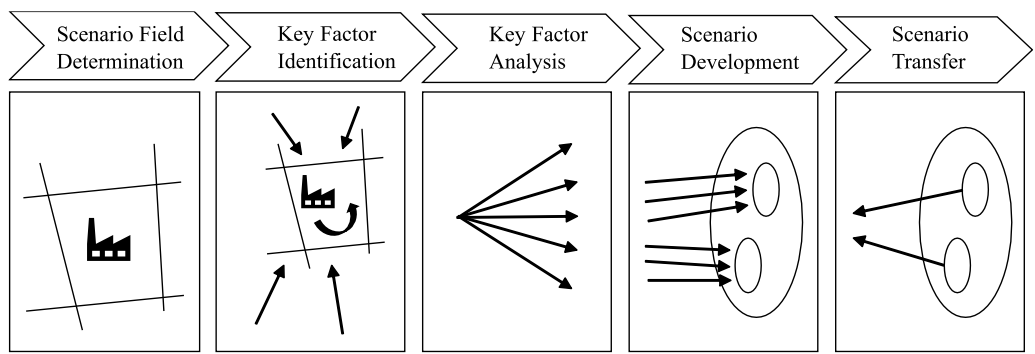

Fig. 1. Process of Scenario Analysis based on (Gausemeier et al., 1998; Schuh et al., 2014; Kosow and León, 2015)

In the first phase of the scenario analysis process, the scenario field is defined. It is determined what the subject is, which aspects are to be considered and which will not be considered. In the second phase, the factors that determine or the scenario field or have an effect on it are described. In the second phase, the factors that determine or influence the scenario field are described. The key factors are variables, parameters, trends, developments and events that are taken into account centrally as the scenario process continues (Kosow and León, 2015). In the third phase, the key factors are analyzed. What possible characteristics or developments are conceivable? In the fourth phase, the scenario development, the possible projections of all key factors are summarized in bundles. Which scenarios are conceivable and how consistent are they? In the last phase, strategies for the decision field are derived.

This approach can be transferred to product development for the purpose of failure analysis. ISO 21448 has even required this since 2022 for the development of road vehicles. A failure can be considered the result of a chain of events which is the definition of a scenario. In order to understand the causality of a failure, it is necessary to examine the connection between causes and results. A formalized description of this sequence is required for a systematic failure analysis. (Iino et al., 2003) uses "actions" as a link between causes and results in a scenario. In (Bielefeld et al., 2020) the "effect" connects cause with result. In the FMEA (failure modes and effects analysis) by (AIAG & VDA, 2022), failure causes lead to a failure mode, which characterizes the malfunctioning behavior. This can lead to different failure effects, which are defined in (AIAG & VDA, 2022) as consequences of a failure. This differs from the definition of (Bielefeld et al., 2020), which is used in the following:

**Definition: An Effect can be understood as a logical, physical or mathematical equation.**

For example, centrifugal force would be an effect according to the definition. It is a function of mass, velocity and radius which can be considered as influencing factors of the scenario. Identifying influencing and key factors in complex product systems pose some challenges.

### 1.3. Challenges in the Identification of Scenario Influencing Factors

Figuring out how to categorize the system elements into cause, action, event, effect, and result, as well as elucidating their interrelationships, is a challenge **(CH1)**. The network of system elements that contribute to a scenario are also called influencing factors (Gausemeier *et al.*, 1998). (Gräßler *et al.*, 2020) compared three approaches of scenario development: intuitive logic approaches (qualitative), cross-impact approaches (quantitative) and consistency-based approaches (both quantitatively and qualitatively). In these cases, the terms "quantitative" or "qualitative" do not refer to the collection of influencing factors, but to the way in which these factors are integrated and further used in the analysis process. Their identification is mostly done with qualitative methods like brainstorming, method 6-3-5 or Delphi method (Gausemeier *et al.*, 1998; Schuh *et al.*, 2014). This poses a challenge **(CH2)** as these methods reach their limits with increasing complexity of the viewed system. In failure analysis, the probability of occurrence and detection of the failure and the severity of the harm must be considered (AIAG & VDA, 2022). Cross-impact approaches consider the probability of occurrence of events. However, the probability of detection and the severity of harm are not considered in any of these approaches **(CH3)**. The identification of failure scenarios and their influencing factors, as described in CH1 to CH3, is of central importance, since it forms the basis for failure and risk analysis, e.g. in (ISO, 2018a, 2018b, 2022; AIAG & VDA, 2022).

## 2. State of the Art: Systematic Identification of Influencing Factors for Scenarios

Existing approaches of the systematic data-based identification of influencing factors are analysed in this section. The databases of Web of Science, Springer, IEEE Xplore, Science Direct and Wiley were searched with the keywords (determining OR determine OR identify OR identification) AND ("influencing factor" OR "key factor"). Articles that describe an approach for data-based identification of influencing factors regarding target values of technical products and their development were considered.

### 2.1. Approaches of systematic identification of influencing factors

Before analysing the influencing factors, they must be collected and pre-processed. This can be done with databases. (Mahboob *et al.*, 2017) divide the scenario in an actor, system and environment model with its own databases for storing the influencing factors. Here, they are collected through the observation of an VR environment. (Bielefeld *et al.*, 2020) use databases for effects, events, the product system and the environment. The selection of influencing factors is not be discussed in more detail. (Yan *et al.*, 2020; Yuan, 2022; Munikhah and Ramdhani, 2022) conducted a literature review to collect influencing factors. (Li *et al.*, 2022) combined multiple sources of data with machine learning to derive influencing factors. For instance, the sources Li et al. used are open street maps, ride sharing trajectory data, number of facilities and user review data. In (Kisenasamy *et al.*, 2022; Yuan, 2022; Alrawi *et al.*, 2022) influencing factors are obtained from the results of a survey. The influencing factors with regard to vehicle emissions were determined in (Chen *et al.*, 2018) using a dynamometer. A similar approach using a test rig was chosen in (Li *et al.*, 2021) to determine factors influencing the performance of an air curtain. (Morais *et al.*, 2022) analysed accident report datasets with a machine learning classification scheme that identifies the relation between human errors and their influencing factors. The categories "organisational", "technological", "individual" and "human execution" of the factors are predefined. (Gräßler *et al.*, 2020) provide an agile approach for scenario modelling. Influencing factors are selected from a database of projects in a defined topic area and supplemented by project-specific factors in an iterative way.

### 2.2. Approaches of determining the interrelationship and interaction of influencing factors

The literature presents different approaches for investigating the network of influencing factors of a scenario that characterizes the connection between causes and results (Fig. 1). (Mahboob et al., 2017) use SysML behaviour and structure models, like internal block diagrams, state machine diagrams or activity diagrams, in order to model the relationship of influencing factors. (Bielefeld et al., 2020) applied a model-based systems engineering approach, the Demand-Complaint-Design (DeCoDe) by (Mistler et al., 2021) for modelling failure networks. DeCoDe defines relations between system elements that can used to draw connections between influencing factors. For instance, in DeCoDe functions fulfill requirements. Components realize functions. Processes use components as resources and realize functions. Persons uses and realizes processes and components and persons realize functions. In (AIAG & VDA, 2022) parameter diagrams are proposed. The diagram includes a model of the product, its inputs and outputs, its functions and requirements and noise factors, which are similar to the scenario influencing factors. In addition, Failure Networks are addressed without going in more detail. The observation of predefined influencing factors by a camera which is monitoring an artificial test track was done by (Yan et al., 2020; Ye et al., 2016). (Li et al., 2022) analysed the correlation between the built environment factors and the multidimensional urban vitality on the street using a multiple regression model. In (Munikhah and Ramdhani, 2022), the interaction between influencing factors is determined by experts with the use of a pairwise comparison. (Gausemeier *et al.*, 1998) project possible developments of each influencing factor in bundles that are then subjected to a consistency check in order to free them from contradictions. (Schuh *et al.*, 2014) applies a SARIMA regression model to analyse the contributing factors.

## 3. Research Questions and Objective

The literature shows that there is a wide range of scenario analysis methods. Scenario-based methods are already established in the strategic planning of products and companies. In the area of technical product development, the methods are becoming increasingly important. However, there are still questions to be answered. A central step in all scenario analysis methods is the analysis of influencing factors and the identification of key factors. This step frequently depends on qualitative methodologies, which lack tailored considerations for the complex structure of causal failure relationships in technical products that interact with their environment. This introduces a level of uncertainty, particularly when applied to complex systems. It must be ensured that the identified factors comprehensively cover the entirety of potential malfunctions that a product may experience throughout its utilization phase. The objective of this research lies in addressing these uncertainties with a robust methodology that adequately considers the complex relationship between the product and environmental systems within its operational contexts. The following research questions were posed to guide the development of the method.

- **RQ 1.** What elements or influencing factors must the chain of events contain in order to describe failures in technical products? How can these factors be systematically identified?
- **RQ 2.** How can actual information from failure databases be used to make the method robust?
- **RQ 3.** How can the relationships between the factors that lead to failures be characterized in order to prepare a further analysis?

The developed method is presented in section 4 and validated in a case study in section 5. A discussion of the results is given in section 6. A conclusion is drawn in section 7.

## 4. Methodology

### 4.1. Elements of the chain of events in scenario-based failure analysis

Section 2 showed that **causes, results, actions and effects** are necessary to describe a chain of events. With regard to failure analysis, the results are limited to failures. Failures are definded by the non-fulfillment of requirements (ISO, 2015). To further narrow the focus, only safety-relevant requirements are considered. A systematic identification of these elements can be accomplished through systems thinking (Haberfellner *et al.*, 2019). Systems consist of interconnected elements that are separated from the environment by a system boundary. A system can be part of a system of systems, allowing the environment to be described. It is essential for a system to fulfill a function by transforming inputs into outputs. The elements of a socio-technical system can be described by components, functions, processes, requirements and persons (Mistler *et al.*, 2021). To examine how inputs, outputs and the environment can be described in more detail in the context of scenario-based failure analysis, reference can be made to the FMEA handbook of (AIAG & VDA, 2022), which defines parameters to be taken into account for product and process FMEAs. Processes are also relevant here, as usage processes describe the use of the product by the user in their environment. In addition to the parameters already described, the FMEA handbook also references control and noise factors. A control factor intentionally adjusted to manipulate the output of product and control its function. A noise factor in a product system represents an uncontrollable variable or external influence that may introduce unwanted variations or disruptions in the product's performance, that can be used specifically to describe the environment. The parameters derived for describing a chain of events in failure analysis of technical products are summarized in Figure 2.

Components, functions and their relationships are part of the system. They can be direct causes of failures or contribute to actions or effects which lead in further steps to failures. Control factors and noise factors are inputs to the system. They can be classified as causes or as inputs into effects. Actions include users. In combination with effects, they link causes with results. As a result, only safety-relevant failures are taken into account, which can be also considered as harm.

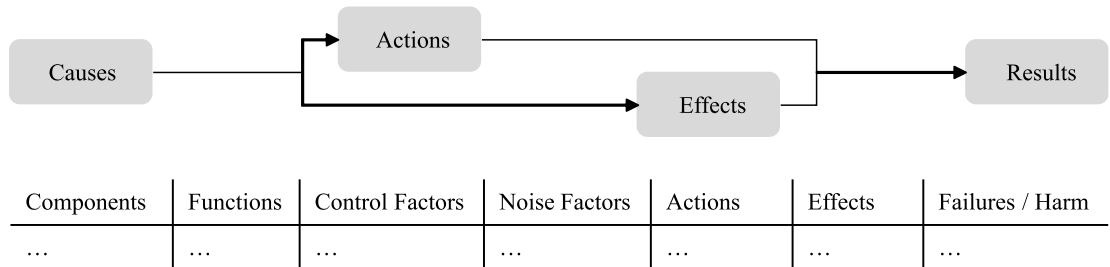

| Components | Functions | Control Factors | Noise Factors | Actions | Effects | Failures / Harm |
|---|---|---|---|---|---|---|
| … | … | … | … | … | … | … |

Fig. 2. Parameters of the chain of events for scenario-based failure analysis

### 4.2. Databases as Sources of Failure Information

A systematic identification of influencing factors can be done by listing the factors that have actually led to failures in the usage of the product. Using a large amount of data ensures the completeness of the analysis. However, new phenomena that have not yet occurred are not examined here. They need to be analyzed in a further step. Databases with a large amount of information include customer review platforms such as Amazon or Ebay. The quality and validity of the information in such platforms varies greatly. This inadequacy could be compensated by a very large amount of data. To filter this data, further steps would be necessary that would go beyond the scope. In order to be able to present the method in a compact manner, fewer but higher quality data should be used. This kind of data can be found in the Rapex database. The Rapex (Rapid Exchange of Information System) database is a European Union system that enables quick information exchange about non-food products posing health and safety risks to consumers. It facilitates coordinated actions among EU member states to remove or restrict such products from the market. Safety gate alerts can be searched on their website (https://ec.europa.eu/safety-gate-alerts/screen/search?resetSearch=true), in which safety-relevant failures of products are described. This information is used in the following.

### 4.3. Failure Sequence Diagram

The Rapex database provides an amount of failure descriptions of given product types. The failure description in text form needs to be transformed in a sequence of events to represent a scenario. The parameters defined in Figure 2 are used for this. From a starting point, which is given in the failure description, for example a component which is not working, a sequence of components, functions, control and noise factors, actions and effects is derived which lead to a failure. The sequence consists of a linear temporal chain of events, in which a single event can only happen at the same time step. Often the underlying effects are not mentioned. These must be identified themselves in order to clarify the causes of events. The Triz effects database or Koller's effects catalog can be used for this purpose (Oxford Creativity, 2022; Koller and Kastrup, 1998). In addition to the effects, the parameters that serve as input for the effects are also listed here. This allows further inputs, e.g. from the environment, to be identified that influences the system. A failure sequence describes an individual scenario of how a product can fail. This single failure can be considered as failure case. In order to cover the entire probability space, as many failure sequences as possible from as many Rapex messages as possible must be documented. This information can therefore be combined into a network of scenarios to more accurately predict the potential failure behavior of the product in the usage. An example of a failure sequence diagram can be found in Figure 4.

### 4.4. Influence Factor Relationship matrix

Once individual failure scenarios have been documented, they must be interconnected. For this relationship matrices can be used. Again, the parameters from Figure 2 are used and are plotted against each other in a cross comparison. The relationship matrices are based on the failure sequence diagrams. The Interrelationships of all failure sequence diagrams are displayed in a single relationship matrix. Because the failure sequence diagrams only allow a single event to occur at a time step, the events are related to each other from a time step to another. In the reference matrix, the related events are linked from the column vector in the direction of the row vector with a "1". The sequence diagrams are displayed step by step pair-wise for each time step in the matrix. If new events occur, the header row and column of the matrix must be added. If the same relationship occurs twice, a "1" is added to the existing "1". The relationship then has the value "2". This describes a weighting of the relationships and elements. The more frequently an element and a relationship occurs and contributes to a failure, the greater is its value. Then the values of the events per row and per column are added together. The added values in the right-hand column describe the active sum per event. It indicates how much one event has influenced another. The added values in the bottom row describe the passive sum per event. It indicates how of one event was influenced by another. These parameters help to improve understanding of the network of influencing factors and their relationship to each other. An example of an influence factor relationship matrix can be found in Figure 5.

### 4.5. Active and passive sum diagram

The visualization of the previously determined passive and active sums per factor is done in the active and passive sums diagram. The y axis of the diagram shows the active sums, while the passive sums are plotted on the x axis. Both axes are normalized between 0 and 100. Each factor is thus given its own unique coordinating. Areas are entered in the diagram that divide the factors into groups. The area near the *y* axis describes dominant factors, which mainly influence other factors. The area near the *x* axis describes reactive factors, which mainly are influenced by other factors. The area between the two axes describes dynamic factors that influence and are influenced in a balanced manner. Key factors that stand out from the other factors in their sums can now be read from the diagram. It should be noted that only statements about frequency can be made here. Information about the influence on the severity of damage, for example, cannot yet be read off here. However, the passive and active sum diagram provides important information for further scenario-based failure analysis. An example can be found in Figure 6.

### 5. Case Study

The methodology is demonstrated on the case study of a hairdryer. 41 Rapex safety alerts were used for the analysis. An example of a hair dryer safety alert, which was used, is shown in Figure 3.

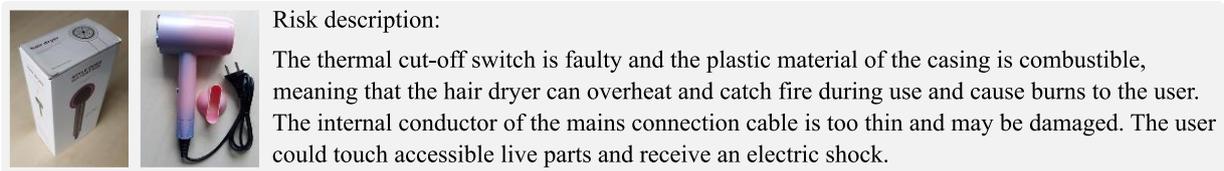

Risk description:

The thermal cut-off switch is faulty and the plastic material of the casing is combustible, meaning that the hair dryer can overheat and catch fire during use and cause burns to the user. The internal conductor of the mains connection cable is too thin and may be damaged. The user could touch accessible live parts and receive an electric shock.

Fig. 3. Risk of burn, electric shock and fire of a hair dryer from Rapex safety gate alert number A12/02261/23 (European Commission,

The safety gate alert describes how the hair dryer shown above could fail in a way that could result in burns, electric shock, or fire, which are three different failure cases which have to be documented in three failure sequence diagrams. The diagram for the case burn is shown in Figure 4.

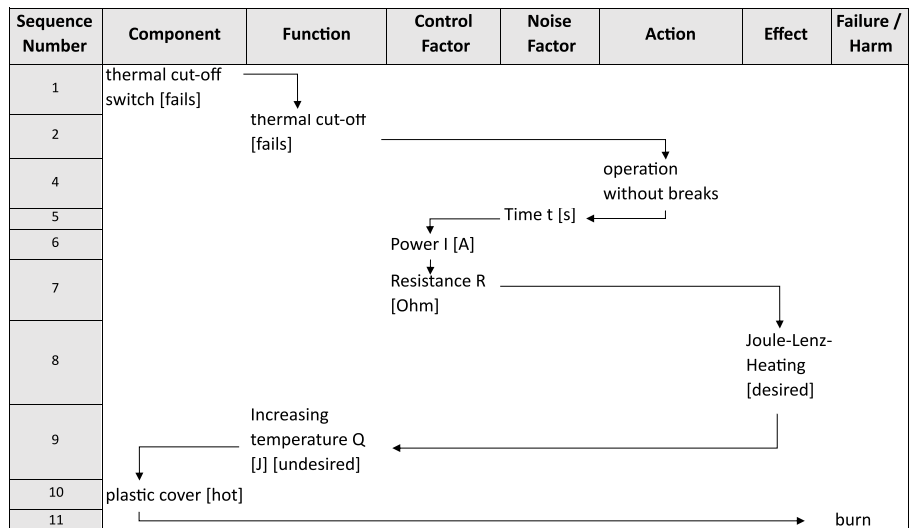

Fig. 4. Failure sequence diagram for the failure of a hair dryer which results in burns derived from the safety gate alert A12/02261/23.

It is visible that the harm "burn" only becomes effective through an effect and an action. The Joule-Lenz-heating effect converts a current through an electrical resistance into heat. However, this only leads to a burn if the user operates the hairdryer without interruption. The Figure shows a single safety case of a Rapex alert. Using this procedure, failure sequence diagrams of 41 safety cases were created and converted into the relationship diagram. An excerpt of the diagram is shown in Figure 5.

| | | | Preventing access | Increasing temperature Q [J] | Control Factor Power I [A] | Force F [N] | ... | active sum | ranking |
|---|---|---|---|---|---|---|---|---|---|
| **Component** | thermal cut-off switch | ... | | | | | ... | 10 | 8 |
| | earth wire | ... | | | | | ... | 1 | 34 |
| | power supply plug | ... | 2 | | | | ... | 6 | 20 |
| | solders | ... | | | | | ... | 1 | 34 |
| | electronic unit | ... | | | 1 | | ... | 1 | 34 |
| | power supply cable | ... | | | | | ... | 7 | 17 |
| | Accessory | ... | | | | | ... | 1 | 34 |
| | heating element | ... | 2 | 1 | | | ... | 6 | 20 |
| | cable jacket | ... | | | | | ... | 2 | 30 |
| | internal conductor of | ... | | | | | ... | 10 | 8 |
| | Thermal Isolation | ... | | 1 | | | ... | 4 | 26 |
| | hair dryer | ... | 2 | 1 | | | ... | 14 | 4 |
| | protective grille | ... | 3 | | | | ... | 8 | 15 |
| ⋮ | ⋮ | ⋮ | ⋮ | ⋮ | ⋮ | ⋮ | | | |
| **passive sum** | | | 11 | 20 | 13 | 1 | | | |
| **ranking** | | | 8 | 3 | 6 | 31 | | | |

Fig. 5. Influence factor relationship diagram (excerpt) of 41 Rapex safety gate alerts of hair dryers

The overall diagram consists of a 50 x 50 matrix in which 46 influencing factors are included, which are sorted into the parameters components, functions, control and noise factors, actions, effects and failures derived in section 4.1. The respective active and passive totals of the rows and columns with the corresponding ranks in relation to all influencing factors can be seen in Figure 5. An overview of all influencing factors determined with the corresponding sums is given in Table 1.

Table 1. Influencing Factors with active and passive sums

| | ID | Influencing Factor | Active Sum | Normalized | Ranking | Passive Sum | Normalized | Ranking |
|---|---|---|---|---|---|---|---|---|
| **Component** | 1 | thermal cut-off switch | 10 | 43,5 | 8 | 0 | 0 | 41 |
| | 2 | earth wire | 1 | 4,3 | 34 | 1 | 4,2 | 31 |
| | 3 | power supply plug | 6 | 26,1 | 20 | 0 | 0 | 41 |
| | 4 | solders | 1 | 4,3 | 34 | 0 | 0 | 41 |
| | 5 | electronic unit | 1 | 4,3 | 34 | 1 | 4,2 | 31 |
| | 6 | power supply cable | 7 | 30,4 | 17 | 0 | 0 | 41 |
| | 7 | accessory | 1 | 4,3 | 34 | 0 | 0 | 41 |
| | 8 | heating element | 6 | 26,1 | 20 | 0 | 0 | 41 |
| | 9 | cable jacket | 2 | 8,7 | 30 | 2 | 8,3 | 28 |
| | 10 | internal conductor of the mains connection cable | 10 | 43,5 | 8 | 8 | 33,3 | 15 |
| | 11 | Thermal Isolation | 4 | 17,4 | 26 | 2 | 8,3 | 28 |
| | 12 | hair dryer | 14 | 60,9 | 4 | 9 | 37,5 | 12 |
| | 13 | protective grille | 8 | 34,8 | 15 | 3 | 12,5 | 25 |
| | 14 | plastic cover | 11 | 47,8 | 7 | 10 | 41,7 | 9 |
| **Function** | 15 | thermal cut-off | 10 | 43,5 | 8 | 10 | 41,7 | 9 |
| | 16 | mechanical strength | 2 | 8,7 | 30 | 1 | 4,2 | 31 |
| | 17 | exchange of accessories | 2 | 8,7 | 30 | 1 | 4,2 | 31 |
| | 18 | spatial separation | 2 | 8,7 | 30 | 2 | 8,3 | 20 |
| | 19 | protection against pulling and twisting cables | 5 | 21,7 | 23 | 5 | 20,8 | 20 |
| | 20 | heat Isolation | 1 | 4,3 | 34 | 1 | 4,2 | 31 |
| | 21 | electrical Isolation | 7 | 30,4 | 17 | 3 | 12,5 | 25 |
| | 22 | preventing access to internal parts | 3 | 13 | 28 | 11 | 45,8 | 8 |
| | 23 | increasing temperature Q [J] | 22 | 95,7 | 2 | 20 | 83,3 | 3 |
| **Control Fac.** | 24 | power I [A] | 15 | 65,2 | 3 | 13 | 54,2 | 6 |
| | 25 | force F [N] | 1 | 4,3 | 34 | 1 | 4,2 | 31 |
| | 26 | voltage U [V] | 5 | 21,7 | 23 | 5 | 20,8 | 20 |
| | 27 | resistance R [Ohm] | 12 | 52,2 | 5 | 14 | 58,3 | 5 |
| **Noise Factor** | 28 | time t [s] | 8 | 34,8 | 15 | 8 | 33,3 | 15 |
| | 29 | concentration of Bis(2-ethylhexyl)phthalat (DEHP) | 1 | 4,3 | 34 | 1 | 4,2 | 31 |
| | 30 | concentration of lead | 1 | 4,3 | 34 | 1 | 4,2 | 31 |
| | 31 | distance between components | 10 | 43,5 | 8 | 10 | 41,7 | 9 |
| | 32 | material thickness d [mm] | 9 | 39,1 | 12 | 9 | 37,5 | 12 |
| | 33 | unstable Fastening | 3 | 13 | 28 | 3 | 12,5 | 25 |

Table 1 (continued). Influencing Factors with active and passive sums

| | ID | Influencing Factor | Active Sum | Normalized | Ranking | Passive Sum | Normalized | Ranking |
|---|---|---|---|---|---|---|---|---|
| Action | 34 | operation without breaks | 9 | 39,1 | 12 | 9 | 37,5 | 12 |
| | 35 | User pulls on the power cable | 4 | 17,4 | 26 | 4 | 16,7 | 24 |
| | 36 | user ignores overheating | 7 | 30,4 | 17 | 7 | 29,2 | 18 |
| | 37 | user touches accessible live parts | 23 | 100 | 1 | 23 | 95,8 | 2 |
| Effect | 38 | Joule-Lenz-Heating | 12 | 52,2 | 5 | 12 | 50 | 7 |
| | 39 | tensile strength | 1 | 4,3 | 34 | 1 | 4,2 | 31 |
| | 40 | Electric Arc | 1 | 4,3 | 34 | 1 | 4,2 | 31 |
| | 41 | Flexural strength | 6 | 26,1 | 20 | 6 | 25 | 19 |
| | 42 | Breakout Strength | 5 | 21,7 | 23 | 5 | 20,8 | 20 |
| | 43 | Electrical Resistance | 9 | 39,1 | 12 | 8 | 33,3 | 15 |
| Harm | 44 | burn | 0 | 0 | 44 | 18 | 75 | 4 |
| | 45 | poisoning | 0 | 0 | 44 | 2 | 8,3 | 28 |
| | 46 | electrical shock | 0 | 0 | 44 | 24 | 100 | 1 |

46 influencing factors lead to a total of 3 types of damage in different scenarios: burns, poisoning and electric shock. Since the harm is defined as the result of the scenario, it only has a passive sum. The active sum corresponds to 0, as no further events are taken into account after the harm occurs. Electric shock is the most, followed by burns. Poisoning has occurred twice. One of them is lead poisoning. The other is poisoning with the chemical DEHP. To derive more information from the summation table, the values are displayed in the active and passive sum diagram (Figure 6).

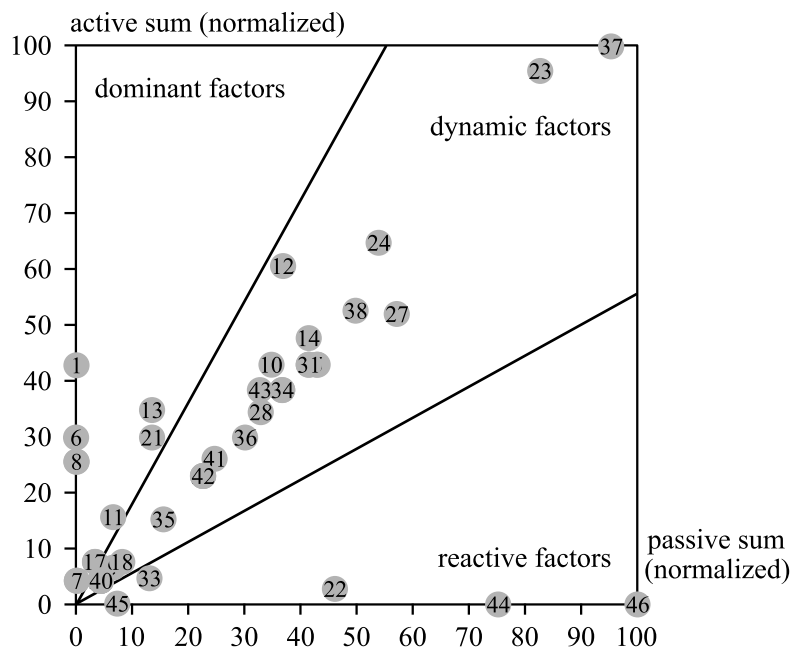

Fig. 6. Active and passive sum diagram of 46 influencing factors

The key factors can now be read from the diagram. The numbers correspond to the IDs of the influencing factors from Table 1. The factors that are recognizable as key factors are the dynamic factors 23 "Increasing temperature" and 27 "Resistance R", the reactive factors 44 "burn" and 46 "electrical shock", which are the results in the scenarios. Also important are the dominant factors 1, 6, 8, 13, 21, the dynamic factors 12, 24, 27, 38 and the reactive factor 22. The labels can be read in Table 1.

## 6. Discussion

The methodology for identifying key factors in failure scenarios shows revealing results. Based on 41 Rapex safety gate alerts, 46 influencing factors were identified that lead to safety-relevant failure regarding the case study example hair dryer. The most important harms are burns and electric shocks, which are also highlighted by the active and passive sum diagram in Figure 6. As expected, the "poisoning" harm was rated as less important because it is based on only two sources. The function of the temperature increase is particularly emphasized when looking at the sums, which is mainly due to the Joule Lenz heating effect. When developing new products, greater attention must therefore be paid to how this effect can be negatively influenced. The current flow, electrical resistance and time play a role. The time factor is particularly difficult to control here, as the influence factor 34 "operation without breaks" also shows. For this purpose, safety devices can be recommended that indicate to the user when the device should be switched off. This should be integrated redundantly to the cut-off switch, as it has repeatedly failed. The damage of the electric shock was often caused by the influencing factor 37 "user touches accessible live parts", was mainly a result of too large distances between components (factor 31) or power cables that were susceptible to damage. During product development, particular attention must be paid to the factors that influence the mentioned risk factors. Secure connections of the power cables, sufficient material thicknesses of the cable insulation and dimensions of the plug fastenings that are matched to the rest of the product should be checked especially in quality assurance tests. The active and passive sum diagram reveals key factors that have higher active and passive sums than the other factors. However, it should be noted that this statement refers to frequencies. Other important parameters of a risk analysis, such as the severity of damage, are not yet considered here. The results mainly show how often an influencing factor has influenced the system and how often it has been influenced itself. Statistical parameters for further failure analysis can be derived from this information. It should also be mentioned that the results of this work only represent the part of the key factors identified by a scenario-based failure analysis. By linking the influencing factors through a relationship matrix, the part of the scenario formation has already been started, which is to be completed by further methods. The use of graphical representation, including failure sequence diagrams and influence factor relationship matrices, aids in visualizing complex interactions among influencing factors. Potential biases in the data, derived from safety databases, may limit the generalizability of identified failure scenarios. The methodology can be extended to various product categories, thereby enhancing product safety across diverse industries. The integration of severity assessments in future iterations could provide a more comprehensive understanding of the impact of identified failure scenarios. Additionally, the potential development of automation tools holds promise for streamlining the analysis process and increasing overall efficiency. There is a risk of over-reliance on empirical data, particularly if safety alerts from existing databases lead to an emphasis on specific types of failures and fail to capture emerging risks.

## 7. Conclusion

In this research article, a data-based methodology for the systematic identification of key factors in safety-relevant failure scenarios was presented. The approach is based on safety gate alerts from the Rapex database, but can also be supplied with failure information from other databases. However, attention must always be paid to the relationship between data quantity and data quality. The methodology serves as a basis for further scenario-based failure analysis. Recommendations for product development have already been derived for the hairdryer case study example. It was shown, that a failure scenario can be described with the parameters "components," "functions", "control and noise factors", "actions", "effects" and "failures / harms" (**RQ1**). It was also shown that linking causes and results in failure chains through the combination of actions and effects is valid. Each of the 41 harms that were analyzed only became effective through an action, an effect or a combination of both. Through the separate documentation in failure sequences, which were later combined in a relationship matrix, information from databases could be used effectively (**RQ2**). Using information from failure databases makes the method robust because it is based on actual failures from the application of the product. However, new failures that have not occurred before cannot be identified this way. For product development, a combination of sources for failure analysis is therefore recommended, which should consist of real information as in the present method, as well as theoretical information that results from classic analysis methods. The value of the presented methodology lies primarily in the provision of a range of information that is produced systematically and covers the significant range of possible failures. The relationship between the influencing factors could be characterized by active and passive sums which were further grouped into dominant, dynamic and reactive factors (**RQ3**). However, these parameters initially only describe the frequency with which the factors occur. For a subsequent risk analysis according to (AIAG & VDA, 2022), the parameters of damage severity and probability of detection still need to be taken into account. Future work will deal with the further process of scenario-based failure analysis of complex product-environmental systems according to Figure 1. In order to optimize the use of information from the databases, it is proposed to support the processes by using tools to automate the output of the diagrams.

**Acknowledgements**

This research is funded by the Deutsche Forschungsgemeinschaft (DFG, German Research Foundation) – Project No. 502859764.

**References**

AIAG & VDA. 2022. Failure Mode and Effects Analysis: FMEA Handbook, Automotive Industry Action Group. Southfield, Michigan, USA.

Alrawi, O.F., Al-Siddiqi, T., Al-Muhannadi, A., Al-Siddiqi, A. and Al-Ghamdi, S.G. 2022. Determining the influencing factors in the residential rooftop solar photovoltaic systems adoption: Evidence from a survey in Qatar. Energy Reports, Vol. 8, pp. 257–26210.1016/j.egyr.2022.01.064.

Amer, M., Daim, T.U. and Jetter, A. 2013. A review of scenario planning. Futures, Vol. 46, pp. 23–40. https://www.doi.org/10.1016/j.futures.2012.10.003.

Bielefeld, O., Loewer, M., Schlueter, N. and Winzer, P. 2020. A new methodology for a model-based and holistic failure analysis for interactions of product and environment by the example of an electrical linear induction motor. In 2020 IEEE International Conference on Systems, Man, and Cybernetics (SMC), Toronto, ON, Canada, IEEE, pp. 4479–448510.1109/SMC42975.2020.9283209.

Chen, L., Wang, Z., Liu, S. and Qu, L. 2018. Using a chassis dynamometer to determine the influencing factors for the emissions of Euro VI vehicles. Transportation Research Part D: Transport and Environment, Vol. 65, pp. 564–573. https://www.doi.org/10.1016/j.trd.2018.09.022.

ElMaraghy, W., ElMaraghy, H., Tomiyama, T. and Monostori, L. 2012. Complexity in engineering design and manufacturing. CIRP Annals, Vol. 61 No. 2, pp. 793–814. https://www.doi.org/10.1016/j.cirp.2012.05.001.

Gausemeier, J., Fink, A. and Schlake, O. 1998. Scenario Management. Technological Forecasting and Social Change, Vol. 59 No. 2, pp. 111–13010.1016/S0040-1625(97)00166-2.

Gräßler, I., Scholle, P. and Thiele, H. 2020. Scenario Technique. In Vajna, S. (Ed.), Integrated Design Engineering, Springer International Publishing, Cham, pp. 615–64510.1007/978-3-030-19357-7_20.

Haberfellner, R., Weck, O.L. de, Fricke, E. and Vössner, S. 2019. Systems Engineering: Fundamentals and Applications. Springer eBook Collection, Birkhäuser, Cham10.1007/978-3-030-13431-0.

Iino, K., Hatamura, Y. and Shimomura, Y. (2003). Scenario Expression for Characterizing Failure Cases. Proceedings of DETC'03, ASME 2003 Design Engineering Technical Conferences and Computers and Information in Engineering Conference Chicago, Illinois, USA.

Intel 2023. Moore's Law. The past, present and future of Gordon Moore's golden rule for the semiconductor industry. Available at: https://www.intel.com/content/www/us/en/newsroom/resources/moores-law.html#gs.449ybn (accessed 21 August 2023).

ISO 2015. ISO 9000:2015: Quality management systems - Fundamentals and vocabulary. International Organization for Standardization, Geneva.

ISO 2018a. ISO 26262-1:2018: Road vehicles - Functional safety - Part 1: Vocabulary. International Organization for Standardization, Geneva.

ISO 2018b. ISO 31000:2018: Risk management. International Organization for Standardization, Geneva.

ISO 2022. ISO 21448:2022: Road vehicles - Safety of the intended functionality. International Organization for Standardization, Geneva.

Kahn, H. and Wiener, A.J. 1967. The Next Thirty-Three Years: A Framework for Speculation. Daedalus, Vol. 96, No. 3, pp. 705–732. http://www.jstor.org/stable/20027066.

Kisenasamy, K., Perumal, S., Raman, V. and Singh, B.S.M. 2022. Influencing factors identification in smart society for insider threat in law enforcement agency using a mixed method approach. International Journal of System Assurance Engineering and Management, Vol. 13 No. S1, pp. 236–251. https://www.doi.org/10.1007/s13198-021-01378-3.

Koller, R. and Kastrup, N. 1998. Prinziplösungen zur Konstruktion technischer Produkte. Springer Berlin Heidelberg, Berlin, Heidelberg10.1007/978-3-642-58755-9.

Kosow, H. and León, C.D. 2015. Die Szenariotechnik als Methode der Experten- und Stakeholdereinbindung. In Niederberger, M. and Wassermann, S. (Eds.), Methoden der Experten- und Stakeholdereinbindung in der sozialwissenschaftlichen Forschung, Springer Fachmedien Wiesbaden, Wiesbaden, pp. 217–24210.1007/978-3-658-01687-6_11.

Li, Q., Cui, C., Liu, F., Wu, Q., Run, Y. and Han, Z. 2022. Multidimensional Urban Vitality on Streets: Spatial Patterns and Influence Factor Identification Using Multisource Urban Data. ISPRS International Journal of Geo-Information, Vol. 11 No. 1, p. 2. https://www.doi.org/10.3390/ijgi11010002.

Li, X., Zhao, X., Jiang, Y., Zhang, M., Wang, L., Liu, Y., Di Xiao, Xu, X., Li, Z. and Wang, Y. 2021. Air curtain dust-collecting technology: Influence factors for air curtain performance. Journal of Wind Engineering and Industrial Aerodynamics, Vol. 218, p. 104780. https://www.doi.org/10.1016/j.jweia.2021.104780.

Mahboob, A., Weber, C., Husung, S., Liebal, A. and Krömker, H. 2017. Model based systems engineering (MBSE) approach for configurable product use-case scenarios in virtual environments. Proceedings of the 21st International Conference on Engineering Design (ICED17), Product, Services and Systems Design, Vol. 3.

Martínez-Cruz, A., Ramírez-Gutiérrez, K.A., Feregrino-Uribe, C. and Morales-Reyes, A. 2021. Security on in-vehicle communication protocols: Issues, challenges, and future research directions. Computer Communications, Vol. 180, pp. 1–20. https://www.doi.org/10.1016/j.comcom.2021.08.027.

Mistler, M., Schlueter, N. and Lower, M. 2021. Analysis of software tools for model-based Generic Systems Engineering for organizations based on e-DeCoDe. In 2021 IEEE International Systems Conference (SysCon), Vancouver, BC, Canada, IEEE, pp. 1–8. https://www.doi.org/10.1109/SysCon48628.2021.9447081.

Moore, G.E. 1998. Cramming More Components Onto Integrated Circuits. Proceedings of the IEEE, Vol. 86 No. 1, pp. 82–85. https://www.doi.org/10.1109/JPROC.1998.658762.

Morais, C., Yung, K.L., Johnson, K., Moura, R., Beer, M. and Patelli, E. 2022. Identification of human errors and influencing factors: A machine learning approach. Safety Science, Vol. 146, p. 105528. https://www.doi.org/10.1016/j.ssci.2021.105528.

Munikhah, I.A.T. and Ramdhani, A.Y. 2022. Capability Factor Identification and Influence Assesment on Supply Chain Resilience in Indonesian Automotive Industry. Jurnal Teknik Industri, Vol. 24 No. 1, pp. 73–82. https://www.doi.org/10.9744/jti.24.1.73-82.

Oxford Creativity. 2022. Triz Effects Database. Available at: https://www.triz.co.uk/triz-effects-database (accessed 29 December 2023).

Schuh, G., Schultze, W., Schiffer, M., Rieger, A., Rudolf, S. and Lehbrink, H. 2014. Scenario-based determination of product feature uncertainties for robust product architectures. Production Engineering, Vol. 8 No. 3, pp. 383–39510.1007/s11740-014-0532-4.

Wartzek, T., Saure, M. and Wilks, C. 2020. Making Future Complexity in Lighting Electronics Manageable. ATZelectronics worldwide, Vol. 15 No. 9, pp. 44–47. https://www.doi.org/10.1007/s38314-020-0249-4.

Yan, X., Chen, J., Bai, H., Wang, T. and Yang, Z. 2020. Influence Factor Analysis of Bicycle Free-Flow Speed for Determining the Design Speeds of Separated Bicycle Lanes. Information, Vol. 11 No. 10, p. 459. https://www.doi.org/10.3390/info11100459.

Yuan, X. 2022. Analysis on the Influencing Factors of College Students' Online Learning during the COVID-19 Epidemic. In 2022 International Conference on Management Engineering, Software Engineering and Service Sciences (ICMSS), Wuhan, China, IEEE, pp. 55–60. https://www.doi.org/10.1109/ICMSS55574.2022.00016.